\documentclass[aps,prl,reprint]{revtex4-1}

\usepackage[dvips]{graphicx}
\usepackage{amssymb,amsfonts,amsmath}
\usepackage{citesort}
\usepackage{color}
\usepackage{soul}

\begin{document}

\title{Giant drag reduction due to interstitial air in sand.}

\author{Tess Homan$^1$ and Devaraj van der Meer$^1$} 

\affiliation{$^{1}$Physics of Fluids Group, Department of Science and Technology,
J.M. Burgers Center, and Mesa+ Institute for Nanotechnology,
University of Twente, P.O. Box 217, 7500 AE Enschede, The Netherlands}


\begin{abstract}
When an object impacts onto a bed of very loose, fine sand, the drag it experiences depends on the ambient pressure in a surprising way: Drag is found to increase significantly with decreasing pressure. We use a modified penetrometer experiment to investigate this effect and directly measure the drag on a sphere as a function of both velocity and pressure. We observe a drag reduction of over 90\% and trace this effect back to the presence of air in the pores between the sand grains. Finally, we construct a model based on the modification of grain-grain interactions that is in full quantitative agreement with the experiments.
\end{abstract}

\maketitle



Every object 
moving through a medium experiences drag \cite{Landau}. 
It is the force that needs to be overcome when an airplane is flying through air, 
a ship is sailing 
the seas, or 
a pole is driven into soil. 
It accounts for most of the energy consumption in all of these examples. 
As a result, even a small 
decrease of drag may cause considerable economic benefit. Strategies that have been employed 
include the careful design of the object's shape \cite{Choi2006} and the addition of polymers 
or bubbles in the surrounding medium \cite{Steinberger2007,White2008,Ceccio2010}, typically leading to a reduction 
of 50\% at most \cite{Sirovich1997,Alben2002,Choi2006b,Rothstein2010,Vakarelski2011,Luu2013}.
Using a modified penetrometer experiment \cite{Stone2004a,Stone2004b,Hill2005,Schroeter2007,Costantino2008,Guillard2013,Matsuyama2014}, 
we show that
air within the pores 
of a loose, fine sand bed is capable of inducing 
drag reduction in excess of 90\%. In particular we prove that this 
reduction accounts for the ambient pressure dependence of drag 
during impact on a sand bed, an observation that puzzled granular scientists for over a decade \cite{Lohse2004,Katsuragi2007,Caballero2007,Royer2007,Goldman2008,Umbanhowar2010,vonKann2010,Clement2011,Royer2011,RuizSuarez2013,Katsuragi2013,Brzinski2013,Nordstrom2014,Joubaud2014,Homan2014b}. 
Finally, we propose a model that fully explains our findings, tracing them back to how air pressure influences the contact forces between the grains.

The reduction of drag on objects moving through a fluid has been an active area of research for decades. 
Traditional methods modify the boundary layer through the addition of polymers \cite{White2008} 
or a modification of the objects' surface \cite{Choi2006} 
and lead to a modest drag reduction of the order of 10-30 percent. Slightly better results may be obtained using more recent strategies which include the injection of (micro)bubbles \cite{Steinberger2007,Ceccio2010} 
and the use of superhydrophobic surfaces \cite{Choi2006b,Rothstein2010}.  
Very recently, a skin friction reduction of over 80\% has been observed in a yield stress fluid (Carbopol) spreading over a rough hydrophobic surface \cite{Luu2013} 
and a record of over 85\% drag reduction was found for a sphere impacting on a liquid by heating it to above the Leidenfrost temperature \cite{Vakarelski2011}. 
Here we show that a similar sphere impacting on a loosely packed bed of fine grains experiences a drag reduction of over 90\%, simply due to the presence of air in the interstitial pores, and explain the reasons why.

It is known that interstitial air may crucially affect the drag an object experiences in a granular material. As a striking example, it may be considerably harder to push a hollow, open-top cylinder into sand than the same one with a closed top. This can be traced back to the release of air below the edges of the container, the so-called blown air effect \cite{Clement2011}. 
A 
similar phenomenon is found in a pre-fluidised granular bed, where the drag on an impacting object is observed to crucially depend on the ambient pressure \cite{Caballero2007,Royer2007,vonKann2010,Royer2011,RuizSuarez2013,Joubaud2014,Homan2014b}. 
Surprisingly, these experiments reveal that the drag on the ball reduces when the air pressure is higher. This results in a deeper penetration for higher 
than for lower ambient pressure.The physical mechanisms behind these phenomena as well as the role played by the air remain largely unknown.   

\paragraph{Experiments} --
To identify the influence of air, we perform a modified penetrometer experiment \cite{Stone2004a,Stone2004b,Hill2005,Schroeter2007,Costantino2008,Guillard2013,Matsuyama2014}, which consists of a sphere connected to a strong linear motor (Figure 1a). The strength and high acceleration of the latter constitutes the difference with traditional penetrometers, such that we can measure beyond the quasi-static regime of very low velocities. The advantage with respect to impact experiments is that we gain full control of the velocity of the penetrating object. The rod that connects the sphere and motor contains a strain gauge which measures the drag force $F$ on the intruder as a function of time (Figure 1b). Combining this curve with the time evolution of the position leads to Figure 1c, where we observe that $F$ increases linearly with the depth $z$ below the surface of the bed. This behaviour is expected from the phenomenological drag law for impact in sand \cite{Lohse2004,Katsuragi2007,Goldman2008,Umbanhowar2010,Katsuragi2013}. 

The experimental setup consists of a $14\times14\times100$ cm$^3$ container partly filled with fine sand (grain size $20-60$ $\mu$m). The sand bed is fluidised with air and subsequently allowed to settle into a very loose packing. 
With a measured bulk density $\rho_s = 0.93$ g/cm$^3$ and a material density  $\rho = 2.21$ g/cm$^3$ this leads to a packing fraction of only 42\%.
To reduce the ambient pressure $P_0$ air can be slowly pumped out of the container, during which 
it was verified that the state of the settled bed is not affected. 
A PVC sphere (diameter $D$) is connected to a rod which runs through a vacuum seal and 
is attached to a linear motor (\textsc{Copley Controls} ServoTube 2506 Module). 
The rod incorporates a load cell (\textsc{Honeywell} model 31) which is located within the container, such that the friction generated in the seal does not contribute to the force measured by the load cell. During each experiment the ball is prescribed to penetrate the sand surface and move down to a certain predefined depth with a constant velocity. While the ball moves down, the forces exerted on the ball are measured simultaneously with the position of the ball. 
The experiments presented here are for a PVC sphere of diameter $D = 3.17$ cm and lead to a maximum measured drag reduction of 88\%. For the largest ball used (perspex, with $D=3.50$ cm) the drag reduction is measured to be as large as 94\%. 

\begin{figure}
\begin{center}
\includegraphics[width=0.7\columnwidth]{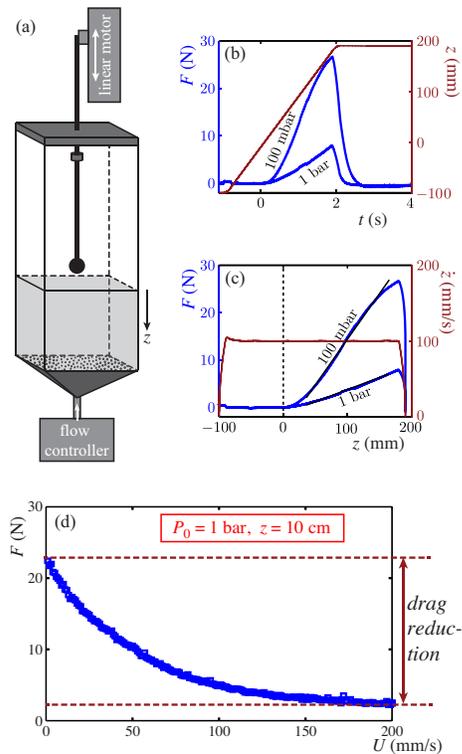}
\caption{Drag force measurements on a sphere penetrating in a sand bed. (a) experimental setup: the linear motor pushes the sphere with constant velocity into the very loose sandbed, while (b) the time evolution of drag force $F$ and vertical position $z$ are measured (for two values of the ambient pressure $P_0$). (c) From this we deduce the depth dependence of the drag, verifying that the velocity $\dot{z}$ has a constant value $U$. (d) By simply increasing $U$ we observe a giant drag reduction inside the sand.}
\end{center}
\end{figure}

\paragraph{Results} --
In the first set of experiments we vary the impact velocity $U=1-200$ mm/s at atmospheric pressure $P_0 = 1.00$ bar and plot the measured drag at a depth of $z = 10.0$ cm below the surface in Figure 1c. 
For the low (quasi-static) value of $U=1$ mm/s where we expect the role of air to be minimal we observe a large drag of $F = 22.5$ N. Increasing $U$ leads to a sharp decrease of $F$ until we reach an approximately constant, plateau value of $F=2.69$ N at the highest attainable velocity $U=200$ m/s. This 
apparently simple observation conceals two remarkable facts. The first is that the drag \emph{decreases} with velocity, whereas in virtually every known situation the drag increases with  $U$ (Stokes', quadratic drag). The second is that there is a drag reduction of 88\%, which for larger spheres than the one used in this experiment (diameter $D=3.17$ cm) 
is observed to become as large as 94\%.

To test whether this drag reduction is due to the interstitial air, we repeat the same experiment at lower values of the ambient pressure $P_0$ (Figure 2a). For low velocities ($U \approx 1$ mm/s), where we expect the influence of air to be minimal, the drag turns out to be largely independent of $P_0$. When we increase $U$, we observe that the lower $P_0$ becomes, the less drag reduction is observed. For the smallest 
value, $P_0=100$ mbar, we find that drag is almost constant leading to a reduction of 20\% at most. These observations imply that indeed the presence of air in the pores between the grains is responsible 
for the drag reduction. This behaviour is corroborated by studying the behaviour of the plateau value of the drag, where in a second set of experiments we fix the velocity to $U = 200$ mm/s and vary the ambient pressure (Figure 2b). Clearly, the plateau value --reported for different depths $z$-- decreases significantly with increasing pressure. Note that the value at $P_0 = 0$ for $z=10$ cm is not taken at $U=200$ mm/s, but is taken from figure 2a and corresponds to the average $F$ measured at the lowest velocity where air should not matter. It smoothly connects with the $z=10$ cm data for $P_0>0$.

\begin{figure}
\includegraphics[width=0.8\columnwidth]{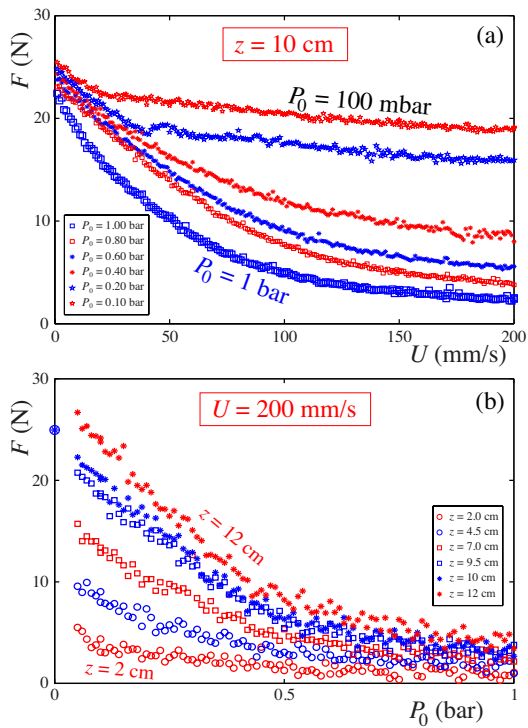}
\caption{Drag force versus velocity and ambient pressure. (a) Drag force $F$ at a depth $z = 10$ cm below the surface as a function of velocity $U$ for different values of the ambient pressure $P_0$. Clearly, for decreasing ambient pressure the drag reduction becomes less pronounced and almost disappears completely for $P_0 = 100$ mbar. (b) The asymptotic value of the drag force $F$ (measured for a large velocity $U = 200$ mm/s) as a function of ambient pressure $P_0$ for different depths $z$ below the surface. For each $z$, $F$ decreases rapidly with pressure.}
\end{figure}

\paragraph{Model and Comparison} --
What is the physical mechanism behind this impressive 
drag reduction? The first thing to realise is that in the absence of air --in the velocity regime studied here-- the drag originates from grains performing work against the typical normal forces between them, either by pushing other grains away or by sliding friction  \cite{Lohse2004,Katsuragi2007,Katsuragi2013}. Both are proportional to the hydrostatic (or lithostatic) pressure $P_h=\rho_s g z$ inside the sand, with $\rho_s$ the bulk density of the sand and $g$ the acceleration of gravity (Figure 3a), i.e., from dimensional analysis one obtains $F=f \rho_s g z D^2$, in which the dimensionless drag coefficient $f$ assumes some numerical value. Let us now assume that the penetrating sphere locally creates an air pressure increase $\Delta P$ around it. Again on dimensional grounds, the air pressure gradient associated with $\Delta P$ gives rise to a reduction of the normal force which should be a function of the ratio $\Pi$ of $\Delta P$ and $P_h$, 
such that the drag force can be written as
\begin{equation}
F \,\,=\,\, \rho_s g z D^2 \,f(\Pi) \,\,=\,\, \rho_s g z D^2 \,\,f\!\left(\frac{\Delta P}{\rho_s g z}\right)\,,
\label{eq_drag}
\end{equation}
where the function $f(\Pi)$ should be a start from a constant value at $\Pi=0$ and afterwards monotonously decrease. 
The precise form of this function $f$ should depend on the details of the sand particle interactions is therefore a priori unknown.

\begin{figure*}
\begin{center}
\includegraphics[width=0.95\textwidth]{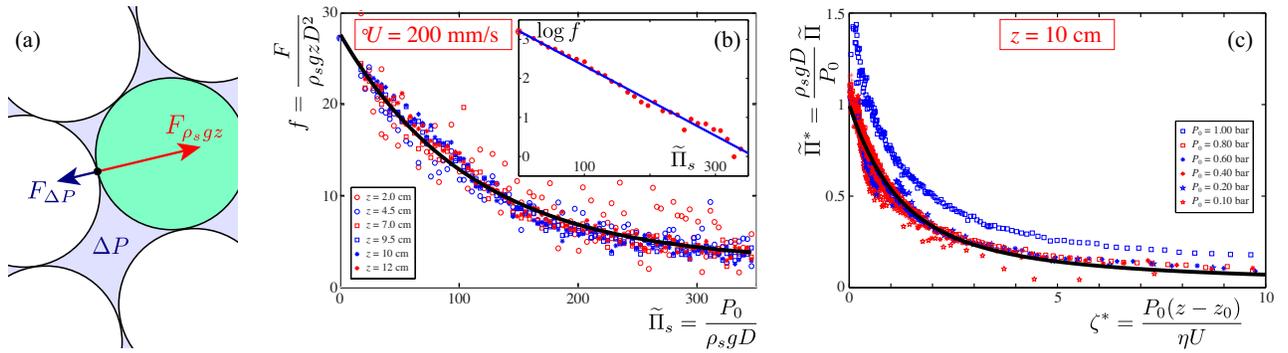}
\caption{Quantitative model for the drag reduction. (a) The physics behind the drag reduction model is the build up of a region with an excess air pressure $\Delta P$ which pushes the particles, which are in a force network determined by the local hydrostatic (or lithostatic) pressure, slightly away from each other thereby decreasing the contact forces. (b) The model provides dimensionless forms of the saturation drag force $f = F/(\rho_s g z D^2)$ and pressure buildup $\widetilde{\Pi} \to \widetilde{\Pi}_s = P_0/(\rho_s g D)$, which leads to a perfect collapse of the data of Figure 2b and to an approximate exponential form for $f(\widetilde{\Pi})$ (inset). (c) The time evolution of the pressure buildup (Eq.~\ref{eq_time_ev}) leads to a second dimensionless form $\widetilde{\Pi}^* = \widetilde{\Pi}\rho_s g D/P_0 = \alpha^{-1}(D/z)(\Delta P/P_0)$ of the pressure difference which is a unique function of the dimensionless depth $\zeta^* = P_0 (z-z_0)/(\eta U)$ with a single fitting parameter $\beta$. Indeed, all but the atmospheric data of Figure 2b collapse onto this function for $\beta = 1.40$ (black line).}
\end{center}
\end{figure*}

The next step is to compute the pressure increase $\Delta P$ as a function of time $t$. First, when the sphere penetrates a sand bed it compacts material in front of it creating an interstitial volume change at a rate $d\Delta V/dt = -\widetilde{\alpha} D^2 U$, with $\widetilde{\alpha}$ some numerical constant. Using isothermal compression (owing to good heat contact between grains and air in the pores) in a pressurised volume $V_0 \propto D^3$ we find from Boyle's law that $d\Delta P/dt = -(P_0/V_0)d\Delta V/dt = \alpha P_0 U/D$, with $\alpha$ a numerical constant.

Secondly, when a pressure $\Delta P$ has built up there is a volume flow rate $Q$ from the pressurised region to the surroundings which is governed by Darcy's law $\vec{Q} = -A\kappa/\mu\vec{\nabla}P$ with $\kappa$ the permeability of the porous medium and $\mu$ the dynamic viscosity of air. Estimating the surface area $A \propto D^2$ and the pressure gradient $\nabla P \propto -\Delta P/D$ we have $Q \propto D\kappa\Delta P/\mu$. Again, $d\Delta P/dt = -(P_0/V_0)Q = -\beta (P_0/\eta)\Delta P$. Here, $\beta$ is a numerical constant and $\eta = \mu D^2/\kappa$ is a constant with the dimension of a dynamic viscosity. The permeability $\kappa$ has been measured independently as $\kappa = 4.0\cdot10^{-12}$ m$^2$ and $\mu = 1.8\cdot10^{-5}$ Pa$\,$s, such that $\eta = 4.5\cdot10^4$ Pa$\,$s. This leads to
\begin{equation}
\frac{d\Delta P}{dt} \,\,=\,\, \alpha\frac{P_0U}{D} \,\,-\,\, \beta \frac{P_0}{\eta}\Delta P\,,
\label{eq_time_ev}
\end{equation}
where $\alpha$ and $\beta$ are numerical constants and $\eta$ is a (fixed) parameter with the dimension of a dynamic viscosity. 

Since $U$ is constant in our experiment Eq.~(\ref{eq_time_ev}) can be solved directly using the initial condition $\Delta P(0) = 0$ \footnote{The use of this initial condition is however somewhat questionable since, if the sphere hits the surface at $t=0$ s, the pressure build-up will start slower than described by Eq.~(\ref{eq_time_ev}) and, also, the initial pressure build-up may be different for different values of $P_0$, due to a difference in the airflow around the object just before impact.} and yields for the slightly modified quantity $\widetilde{\Pi}\equiv \Pi/\alpha = \Delta P/(\alpha \rho_s g z)$  
\begin{equation}
\widetilde{\Pi} \,\,=\,\, \frac{1}{\beta}\frac{\eta U}{\rho_sgDz} \left[ 1 \,\,-\,\, \exp\left( -\beta \frac{P_0}{\eta U}z\right)\right]\,,
\label{eq_sol}
\end{equation}
Note that physically, the introduction of $\widetilde{\Pi}$ expresses that we are not interested in the absolute value of $\Delta P$ (which we cannot measure anyway) but only in establishing its effect on the drag, which is just as well quantified by $\Delta P/\alpha$.

For fixed $z$, if $U$ is sufficiently large, $\zeta \equiv \beta P_0z/(\eta U)$ is small and from first order Taylor expansion of $\exp(-\zeta)$ we find that $\widetilde{\Pi}$ goes to a constant saturation value
\begin{equation}
\widetilde{\Pi} \to \widetilde{\Pi}_s \,\,=\,\, \frac{P_0}{\rho_s g D} \qquad \textrm{for large }U \,,
\label{eq_sat}
\end{equation}
which is independent of both velocity $U$ and depth $z$, in qualitative agreement with the experimental observations (cf. Figs. 1 and 2). 
Moreover, since $\widetilde{\Pi}_s$ is proportional to the ambient pressure, $\Pi_s$ decreases when $P_0$ becomes smaller, leading to a smaller decrease of the drag force [Eq.~(\ref{eq_drag})], just as observed in experiment. In addition, this limit provides us with a way of measuring the unknown function $f(\widetilde{\Pi})$: We measure $f = F/\rho_s g z D^2$ as a function of $P_0$ in the saturation limit of large velocity $U=200$ mm/s and plot the result as a function of the saturation value $\widetilde{\Pi}_s$. This is what is presented in Figure 3b: The data indeed collapse onto a single curve which, incidentally, may be fitted by an exponential $f = f_s + f_0 \exp(-\widetilde{\Pi}/\widetilde{\Pi}_0)$ (see inset). As expected, $f$ is a monotonously decreasing function of $\widetilde{\Pi}$.
  
Finally, we turn to our complete dataset, i.e, including measurements done at smaller values of $U$. For every value of the dimensionless drag force $f = F/\rho_s g z D^2$ we determine $\widetilde{\Pi}$ using the (inverted) functional form determined from Figure 3a. 

We note that, introducing $\widetilde{\Pi}^* \equiv \widetilde{\Pi}\rho_s g D/P_0 = \alpha^{-1}(D/z)(\Delta P/P_0)$, we can rewrite Eq.~(\ref{eq_sol}) more economically as
\begin{equation}
\widetilde{\Pi}^* = \frac{1-\exp(-\beta\zeta^*)}{\beta\zeta^*}
\end{equation}
To compare model and theory we subsequently plot $\widetilde{\Pi}^*$ versus $\zeta^* = \zeta-\zeta_0 = P_0 (z-Ut_0)/(\eta U)$ after allowing for a (ambient pressure dependent) shift $t_0$ in $t$ to correct for ambiguities in properly defining the origin ($t=0$ s, $z=0$ m) in Figure 3c. 
With just one fitting parameter ($\beta=1.40$), the data fits well with the model, especially for the lower values of $\zeta^*$ where the pressure buildup is appreciable.

\paragraph{Conclusion} -- 
In conclusion, we experimentally observed a giant drag reduction of over 95\% inside a bed of very loose, fine sand. By varying the ambient pressure were able to trace this effect back to the presence of air inside the pores between the grains. Secondly, we postulated that the physical mechanism behind this drag reduction is provided by an excess pressure buildup in front of the sphere that diminishes the contact forces between the sand grains by pushing them away from each other. Finally we constructed a model for the pressure buildup and the resulting drag reduction that is in full quantitative agreement with the experimental observations.   

The only case in which the pressure buildup appears to be significantly larger than that expected from the model is that of atmospheric pressure, where $\widetilde{\Pi}^*$ reaches values considerably higher than the expected maximum of $1$. This may well be connected to the fact that there is experimental evidence that the size of the compression region in which the pressure buildup occurs depends on the ambient pressure  \cite{Royer2011,Homan2014b}, which implies that the parameters $\alpha$ and $\beta$ themselves are not constant (as assumed in the model), but in fact functions of the ambient pressure $P_0$.  

Finally, the interplay of the experimentally observed giant drag reduction and the quantitative model is expected to open pathways towards a full control of the drag an object experiences in a granular medium, leading to considerable economic benefit. This may range from taking appropriate measures to stabilise soil to developing strategies for minimising the drag encountered when driving piles into the ground.

\end{document}